\documentclass[twocolumn,preprintnumbers,amsmath,amssymb]{revtex4-2}

\usepackage{graphicx}  
\usepackage{sidecap}
\begin{document}
\title{
Ballistic Injection Terahertz Plasma Instability in Graphene n$^+$-i-n-n$^+$  Field-Effect Transistors and Lateral Diodes}
\author{V. Ryzhii$^{*1,2}$,  M. Ryzhii$^3$, A. Satou$^1$,  V. Mitin$^4$,  M. S. Shur$^5$, and  T.~Otsuji$^1$,}
\address{
$^1$Research Institute of Electrical Communication, Tohoku University, Sendai 980-8577, Japan\\
$^2$Institute of Ultra High Frequency Semiconductor Electronics of RAS,
 Moscow 117105, Russia\\
 $^3$Department of Computer Science and Engineering, University of Aizu, Aizu-Wakamatsu 965-8580, Japan\\
$^4$Department of Electrical Engineering, University at Buffalo, SUNY, Buffalo, New York 14260, USA\\
$^5$Department of Electrical, Computer, and Systems Engineering, Rensselaer Polytechnic Institute, Troy, New York 12180, USA\\
* Corresponding author: e-mail v-ryzhii(at)riec.tohoku.ac.jp, Phone+81-242-222-379
 }
\begin{abstract} 
We analyze the operation of the graphene n$^+$-i-n-n$^+$ field-effect transistors (GFETs) and  lateral diodes (GLDs)  with the injection of ballistic electrons into the n-region. The momentum transfer  of the injected ballistic electrons could lead  to an effective Coulomb drag of 
the quasi-equilibrium electrons in the n-region and  
  the plasma instability in the GFETs and GLDs. The instability enables the generation of terahertz radiation.
 The obtained results can be used for the optimization of the structures under consideration for
different devices, in particular,  terahertz emitters. 
\end{abstract} 

\maketitle

\newpage
\section*{1. Introduction}

The plasma instability and the related self-excitation of plasma oscillations in two dimensional (2D) structures 
can lead to the generation of the terahertz (THz) radiation~\cite{1}. This effect was reported in many theoretical and
experimental papers~\cite{2,3,4,5} (see also the references therein). The 2D structures with the graphene channel (G-channel)
have advantages compared with those based on the standard materials.  
The  high-energy  ballistic electrons (BEs)~\cite{6,7,8,9,10,11,12} injected  into  the G-channel   lead to 
an effective Coulomb drag of the equilibrium carriers~\cite{13,14,15,16,17} strongly affecting the device characteristics.
As demonstrated recently~\cite{18,19},  the Coulomb interaction of the injected BEs with
the quasi-equilibrium electrons (QEs) results in   the dragged electrons (DEs) 
in
the gated G-channel 
of the n$^+$-i-n-n$^+$ graphene field effect transistors (GFETs) and the dragged  holes (DH) in the p$^+$-p-i-n-n$^+$ graphene tunneling transistors (GTTs)~\cite{20}. The drag effect is associated with the linearity of the electron and hole energy-momementum relations in G-channels and followed from  the specifics of  the kinematics of the carrier pair collisions~\cite{21,22,23} and 
can enable the plasma instability associated with the internal current amplification~\cite{14}.  

In this paper, we analyze the high-frequency characteristics of the n$^+$-i-n-n$^+$ G-channel structures
with the gated n-region (called as the GFETs) and with the ungated n-region, to which we refer to as the graphene lateral diodes (GLDs). The Coulomb drag in the  GFETs and GLDs was considered recently~\cite{17}.
We now use the distributed (waveguide) model of the electron plasma in the n-regions instead of
the previous  model based on the equivalent circuits with the lumped capacitance and kinetic inductance of these regions.
The lumped-element model describes well the fundamental plasmonic resonance in the GFETs and GLDs. 
However, the consideration
of the resonance harmonics 
and the analysis of GLDs 
require the distributed model. The electron viscosity can substantially affect the plasma resonances, especially fot higher harmonics. Accounting for the viscosity also requires using the distributive model.
 A  substantial distinction of the plasma oscillations spectra in the gated and ungated channels~\cite{24,25,26,27} leads to markedly different characteristics of the  GFETs and GLDs
despite a similarity of the drag effect in these devices. 
In particular, the plasmonic resonances in the GLDs
can correspond to markedly higher frequencies compared to the GFETs with the same n-region length.

Below we calculate the frequency-dependent impedance of the  GFETs and GLDs as a function of the structural characteristics. 
We demonstrate that the real part of the impedance can be negative in the THz frequency range where 
the impedance imaginary part changes  sign. This corresponds to the plasma instability and the self-excitation 
of THz plasma oscillations~\cite{28} enabling the THz radiation emission using a pertinent antenna.
We calculate the growth rate of the self-excited plasma oscillation as a function of the drag factor and the device structural parameters.

The paper is organized as follows. In Sec.~2, we discuss the GFET and GLD device model describing the BE and QE transport and the role of DEs. In Sec. 3, we derive the spatial distribution of the ac potential in GFETs and calculate their frequency-dependent impedance.
Section~4 deals with the calculation of the GLD frequency-dependent impedance using the 2D Poisson equation for
the spatial distribution of the ac potential. Using the obtained expression for the frequency-dependent impedance,
we analyze the instability of the steady-state current flow in the GFETs (the plasma instability toward the self-excitation of the coupled oscillation of the potential and the electron density)
and find the instability conditions  and the oscillations growth rate.
In Sec.~5, the results of the  distributed device model are compared with those using  a  lumped circuit model. We also comment on the role
of the GLD contacts.
In Sec. 6 we summarize the main conclusions.

\section*{2. Device model}

Figure~1 shows the schematical views of the  GFET and GLD under consideration.
The n-region in the GFET is electrostatically induced by the gate voltage $V_g >0$.
The formation of the n-region in the GDs is due to the remote donor doping. Such a selective doping can provide the electron densities  sufficient for the effective interaction with the BEs injected from the n$^+$-source contact region via the i-region and for the pronounced plasmonic response without marked sacrificing of the QE mobility. 
Similar devices can be implemented using the p$^+$-i-p-p$^+$ structures and
using G-multilayer gated and selectively doped structures with a commensurate improvements in performance.
The  GFETs and GLDs under consideration are forward biased by the source-drain dc voltage $V_0$.
It is assumed that the conditions of the ballistic transport of the injected BEs and the effective transfer of
the BE momentum to the QEs are fulfilled. This primarily implies that the lengths, $l_i$ and $l_n$, of the i- and n-regions, respectively,
are not too large (in the $\mu$m range or somewhat smaller, see~\cite{17,18} for detailes), providing sufficiently large Coulomb drag factor $b$ ($b \gtrsim 1$).

\begin{figure}
\includegraphics[width=7.0cm]{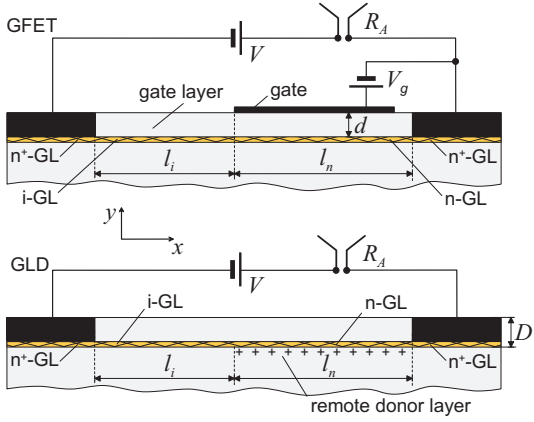}
\caption{Schematic view of   GFET   and  GLD circuit with  an antenna ($R_A$ is the
antenna radiation resistance).
} 
\label{F1}
\end{figure}

Considering  that the net source-drain voltage 
comprises the signal component $V= V_0 + \delta V_{\omega}\exp(-i\omega t)$ with $\delta V_{\omega}$ and $\omega$
being the signal amplitude and frequency, respectively, we obtain:

\begin{eqnarray}\label{eq1}
\delta J_{\omega}^{BE} = \frac{\sigma_i\delta \varphi_{\omega}}{l_i}\biggr|_{x=0}, \qquad  
\delta J_{\omega}^{QE} = \sigma_n
\frac{i\nu_n}{(\omega+ i\nu_n)}\frac{d \delta\varphi_{\omega}}{dx}.\qquad 
\end{eqnarray}
Here the density of the BE ac current 
across
 the i-region ($-l_i \leq x\leq 0$) and  the density of the QE ac current in the  n-region ($0\leq x \leq l_n$), respectively,
$\sigma_i = \kappa\,v_W/2\pi$ is the conductivity of the i-region (disregarding the BE transit-time delay), $\sigma_n = (e^2\mu_n/\pi\hbar^2\nu_n)$ is the n-region Drude dc conductivity,
$\kappa$ is the effective dielectric constant of the media surrounding the G-channel, 
 $v_W\simeq 10^8$~cm/s is the characteristic electron velocity in G-channels, $ \tau_n$ is the QE momentum relaxation time ($\nu_n = \tau_n^{-1}$ represents the frequency of the QE collisions with impurities and acoustic phonons), $\mu_n$ and $\Sigma_n$ are the QE Fermi energy and density with $\mu_n \simeq \hbar\,v_W\sqrt{\pi\Sigma_n}$, $e$ is the electron charge,  $\hbar$ is the Planck constant, and $\delta \varphi_{\omega} = \delta \varphi_{\omega}(x, y)|_{y=0}$ expresses the ac potential spatial  distribution  along the axis $x$  directed in the G-channel plane ($y=0$)  with $x = -l_i$ and $x = l_n$ corresponding to the coordinate of the n$^+$-source and drain region, respectively.   The frequency dependence of the ac conductance reflects the kinetic inductance of the QE system. 

The DE ac current  is given by~\cite{17,18}

\begin{eqnarray}\label{eq2}
\delta J_{\omega}^{DE} = b\Lambda\delta J_{\omega}^{BE}, 
\end{eqnarray}
where $b$ is the drag factor~\cite{16,17}, $b\Lambda = d J_{0}^{DE}/d J_{0}^{BE}$, $J_{0}^{BE} = \sigma_i\Phi_{0}/l_i$ is the dc bias current, and $\Phi_{0}$
is the dc potential at $x= 0$ corresponding to the bias voltage $V_0$.

\section*{3. GFET impedance spectral characteristics}

\begin{figure*}[t] 
\includegraphics[width=11.0cm]{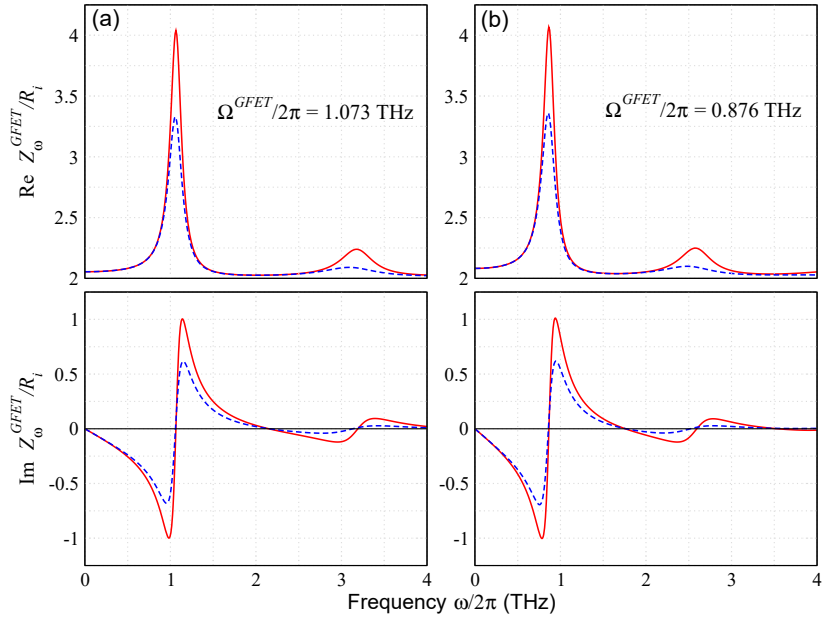}
\caption{Real and imaginary parts of the  GFET impedance, Re~$Z_{\omega}^{GFET}/R_i$ and  Im~$Z_{\omega}^{GFET}/R_i$, versus signal frequency $\omega/2\pi$ at  weak drag effect ($b\Lambda = 0.5$): 
(a) $\Omega^{GFET}/2\pi = 1.073$~THz, $\eta = 9.2$, $\kappa = 4$ and (b) 
$\Omega^{GFET}/2\pi  = 0.876$~THz, $\eta = 6.1$, $\kappa = 6$.
Solid lines correspond to $h = 250$~cm$^2$/s and dashed lines correspond to $h = 500$~cm$^2$/s.
} 
\label{F2ab}
\end{figure*}

\begin{figure*}[t] 
\includegraphics[width=11.0cm]{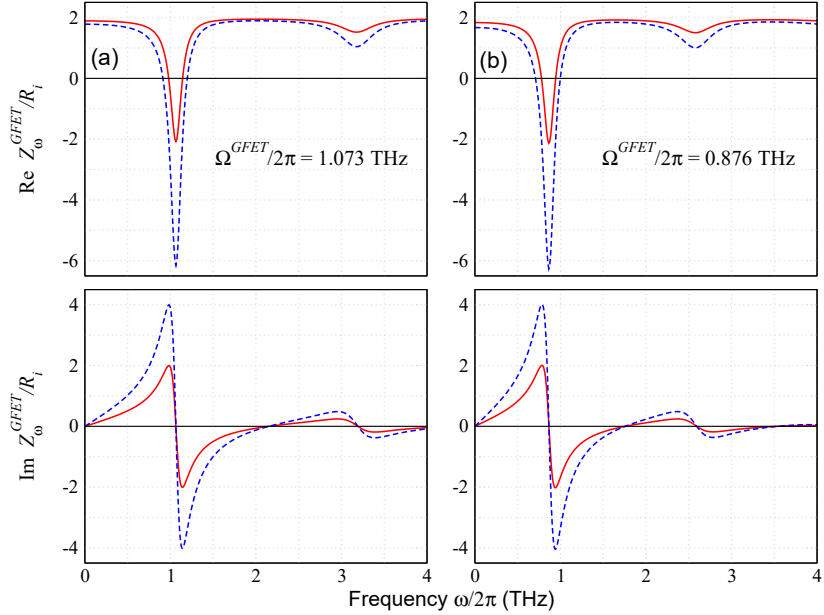}
\caption{Real and imaginary parts of the  GFET impedance, Re~$Z_{\omega}^{GFET}/R_i$ and (b) Im~$Z_{\omega}^{GFET}/R_i$, versus signal frequency $\omega/2\pi$ for GFETs with (a)
$\Omega^{GFET}/2\pi = 1.073$~THz,  and (b) $\Omega^{GFET}/2\pi = 0.876$~THz  
($h = 250$~cm$^2$/s,  $b\Lambda = 2.0$ - solid lines and   
$b\Lambda = 3.0$ - dashed lines).
} 
\label{F3ab}
\end{figure*}

The standard procedure of reducing of the general system of electron plasma hydrodynamic equations coupled with the 2D Poisson equation~\cite{29,30,31,32}
using in the gradual channel approximation~\cite{33} yields  the 
 spatial distribution of the ac potential $\delta \varphi_{\omega} =\delta \psi_{\omega}(x, y)|_{y=0}$, where  
$\delta \psi(x,y)$ is the potential around the n-region of the G-channel ($y = 0$), by the following equation
(see, for example,~\cite{29}):

\begin{eqnarray}\label{eq3}
\frac{d^2 \delta \varphi_{\omega}}                                                                                                                                                                                                                                                                                                                                                                                                                                                                                                                                                                                                                                                                                                                                                                                                                                  
{ d x^2} +\frac{\omega(\omega+i\nu_n)}{s^2} (\delta \varphi_{\omega} -\delta V_{\omega})= 0
\end{eqnarray}
with the boundary conditions at the edges of the n-region: $\delta \varphi_{\omega}|_{x=0} = l_i\delta J_{\omega}^{BE}/\sigma_i$ and 
$\delta \varphi_{\omega}|_{x=l_n} = \delta V_{\omega}$, 
$\delta J_{\omega}^{BE}$
coincides with the net ac terminal  current.
These boundary conditions are valid for all devices under consideration.
 The plasma velocity in the gated n-region is given by $s = \sqrt{4e^2 \mu_nd/\kappa\hbar^2}$ with $d \ll l_n$  being the thickness  of the layer separating the gate and the n-region of the channel.
In reality, the electron fluid viscosity (see also~\cite{34,35,36}) 
can strongly affect the electron dynamics, 
resulting  in a strong damping of the plasma resonances, particularly, affecting the higher resonances, their height and widths~\cite{1,34}.  
This is 
due to the deviation of the ac potential and electron density spatial distributions from a linear function.
To account for the viscosity effect, we put in Eq.~(3) $\nu_n= {\overline \nu}_n
+ h k^2$~\cite{34}, where  $h$ is the QE viscosity and $k = \omega/s$ is the plasma wave number. Considering that for the characteristic plasma frequency $\Omega^{GFET}$ one obtains    
$\Omega^{GFET}/s = \pi/2\l_n$, we set $\nu_n= {\overline \nu}_n
+ h (\pi^2/4l_n^2)(\omega/\Omega^{GFET})^2$. It is worth mention that both the electron drag and the elevated electron viscosity in G-channels originate from the specifics of the Coulomb interaction of the 2D electrons with the linear dispersion law.

Solving Eq.~(3) with the boundary conditions under consideration, we obtain

\begin{eqnarray}\label{eq4}
\delta \varphi_{\omega} - \delta V_{\omega}= 
(\delta \varphi_{\omega}|_{x=0} - \delta V_{\omega})
  \biggl[\cos \biggl(\frac{\gamma_{\omega}^{GFET}x}{l_n}\biggr)\nonumber\\
- \frac{\cos(\gamma_{\omega}^{GFET})}{\sin(\gamma_{\omega}^{GFET})}\sin\biggl(\frac{\gamma_{\omega}^{GFET}x}{l_n}\biggr)\biggr].
\end{eqnarray}
In Eq.~(4),

\begin{eqnarray}\label{eq5}
\gamma_{\omega}^{GFET} = \frac{\pi\sqrt{\omega(\omega + i\nu_n)}}{2\Omega^{GFET}},
\end{eqnarray}
where the plasma frequency is given by

\begin{eqnarray}\label{eq6}
 \Omega^{GFET} = \frac{e}{\hbar}\sqrt{\frac{\pi^2\mu_nd}{\kappa\,l_n^2}}  \propto \sqrt{\frac{d}{l_n^2}}.
\end{eqnarray}

Substituting $\delta \varphi_{\omega}$ from Eq.~(4) into Eq.~(1), equalizing $\delta J_{\omega}^{BE}$ and  $(\delta J_{\omega}^{DE} + \delta J_{\omega}^{QE}|_{x =0})$ (i.e., using the Kirchhoff's circuit law),  
 and accounting for the antenna radiation resistance $R_A$, we arrive at the following expression for the  GFET impedances $Z_{\omega}^{GFET}
= \delta V_{\omega}/H\delta J_{\omega} + R_A$:

\begin{eqnarray}\label{eq7}
\frac{Z_{\omega}^{GFET}}{R_i} = -i\frac{(1 -b\Lambda)}{\eta}\frac{(\omega + i\nu_n)}{\nu_n}\frac{\tan (\gamma_{\omega}^{GFET})}{\gamma_{\omega}^{GFET}} + 1 + \rho_A.\qquad
\end{eqnarray}
Here $H$ is the device size in the $z$-direction perpendicular to
the source-drain current direction (device width), $R_i =\l_i/H\sigma_i$ is the i-region resistance,  $\rho_A =R_A/R_i$, 
$\eta = R_i/R_n$ is the ratio of the n-region and i-region resistances with
$R_n = l_n/H\sigma_n$, so that $\eta = l_i\sigma_n/l_l\sigma_i$. 
In the absence of the electron drag at $\omega$ tending to zero, $Z_{\omega}^{GFET}$ tends to 
$Z_0^{GFET} = R_n + R_i + R_A$.

Using the properties of the trigonometric functions~\cite{37}, the GFET impedance given by Eq.~(7) can be presented in the form explicitly
expressing its resonant behavior:
 
\begin{eqnarray}\label{eq8}
\frac{Z_{\omega}^{GFET}}{R_i} = -i\frac{8}{\pi^2}\frac{(1 -b\Lambda)}{\eta}\frac{(\omega + i\nu_n)}{\nu_n}\nonumber\\
\times
\sum_{m=1}^{\infty}\frac{(\Omega^{GFET})^2}{(2m-1)^2(\Omega^{GFET})^2 - \omega(\omega + i\nu_n)} 
+ 1 + \rho_A.
\end{eqnarray}

As follows from Eq.~(7) and more clearly seen from Eq.~(8), when $\Omega^{GFET} \gg \nu_n$,
the real part of the GFET impedance Re~$Z_{\omega}^{GFET}$ as a function of the signal frequency $\omega$ exhibits the resonant peaks.
The resonant frequencies $\omega_{2m-1}$, where  $m = 1,2,3,...$ is the resonance index,  are given by
 
\begin{eqnarray}\label{eq9}
\omega_{2m-1} \simeq (2m-1)\sqrt{(\Omega^{GFET})^2 - \nu_n^2}.
\end{eqnarray}
For the height of the resonant peak we obtain

\begin{eqnarray}\label{eq10}
\biggl|\frac{{\rm Re} Z_{\omega}^{GFET}}{R_i}\biggr| =\frac{8}{\pi^2}\frac{|1-b\Lambda|}{\eta}\biggl(\frac{\Omega^{GFET}}{\nu_n}\biggr)^2
+ 1 + \rho_A.\qquad
\end{eqnarray}
The resonant peaks are pronounced if $(\Omega^{GFET}/\nu_n)^2 \gg \eta = R_i/R_n$. Since $\Omega^{GFET} \propto l_n^{-1}$ and $R_n \propto l_n\nu_n$, the latter inequality can be satisfied when the product $l_n\nu_n$ is sufficiently small.
A marked increase in $\nu_n$ with increasing peaks index $m$ associated with the viscosity effect, results
in relatively small height of these peaks.
Assuming that $h =250 - 500$~cm$^2$/s (see the estimates for the electron viscosity~\cite{34})
and $l_n = 0.5~\mu$m, for the quantity 
${\tilde \nu}_n = h(\pi^2/4l_n^2) \simeq 0.25 - 0.5$~ps$^{-1}$.
Using Eq.~(8), at ${\overline \nu}_n = 0.75$~ps$^{-1} $ $h =250$~cm$^2$/s for the ratio of the fundamental ($m = 1$) peak height and the next peak ($m=2$) height
we obtain $\sim 9$.

Figure~2 shows the spectral characteristics of  the normalized real and imaginary parts of the GFET impedance, Re~$Z_{\omega}^{GFET}/R_i$ and  Im~$Z_{\omega}^{GFET}/R_i$, calculated using Eq.~(7) for different values of the plasma frequencies
$\Omega^{GFET}$ and  the electron viscosity $h$.
We assumed that $\mu_n =25$~meV, $l_n = 0.5 \mu$m, $l_n/\l_i = 5$,  $d =0.05~\mu$m,
$\kappa = 4$ and  6, $\eta =6.1$ and  - 9.2,  ${\overline \nu}_n = 0.75$~ps$^{-1}$, $h = (250 - 500)$~cm$^2$/s, $\rho_A = 1$,
and 
$b\Lambda = 0.5$. The latter parameters  correspond to $b= 0.25$ and $J_0^{BE} = 1.41$~A/cm and can be related to 
the GFETs at room temperature~\cite{17}.
One can see from Fig.~2(a)
that the impedance real part exhibits a series of the markedly damping resonant peaks.

Figure~3 shows thereal and imaginary parts of the GFET frequency-dependent impedance normalized by $R_i$ under the  condition of relatively strong drag effect ($b\Lambda > 1$). Other parameters are the same as for Fig.~2. 
One can see that at chosen parameters near the plasma resonances (fundamental) with the frequencies $\omega_1/2\pi =\Omega^{GFET}/2\pi = 1.073$~THz and $\omega_1/2\pi =\Omega^{GFET}/2\pi = 0.876$~THz
 the GFET impedance is negative.
The impedance  imaginary part changes  sign at  the resonant frequency, i.e., in the range where Re $Z_{\omega}^{GFET} < 0$ (around of the above resonant frequencies). However, the latter takes place only for the fundamental plasma resonance due to a strong damping associated with the viscosity.

As pointed out previously~\cite{18,19,20}, such a situation implies the possibility of the plasma instability (see below), i.e., the self-excitation of the plasma oscillations (see, for example,~\cite{28}).

\begin{figure*}[t] 
\includegraphics[width=11.0cm]{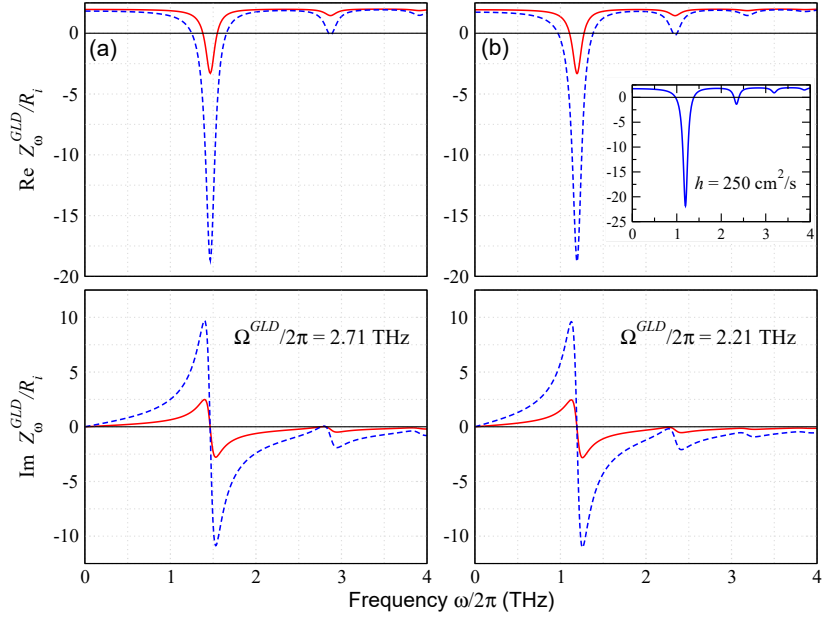}
\caption{Real and imaginary parts of GLD impedance, Re~$Z_{\omega}^{GLD}/R_i$ and  Im~$Z_{\omega}^{GLD}/R_i$, versus signal frequency $\omega/2\pi$ for GLDs with (a)
$\Omega^{GLD}/2\pi = 2.71$~THz, $\eta = 4.6$  and (b) $\Omega^{GLD}/2\pi = 2.21$~THz, $\eta = 3.05$ 
($h = 1000$~cm$^2$/s,  $b\Lambda = 2.0$ --solid lines and   
$b\Lambda = 3.0$ - dashed lines). Inset shows the  Re~$Z_{\omega}^{GLD}/r_i$ versus frequency dependence for the same parameters as in  panel (b), but with smaller viscosity ($h = 250$~cm$^2$/s).
} 
\label{F4ab}
\end{figure*}

\section*{4. Two-dimensional potential distribution  in GLD and its impedance}
The potential distribution $\delta \psi_{\omega}(x,y)$ around the n-layer in the GLD is governed by the 2D Poisson equation in the following form~\cite{29,30,31,32}:

\begin{eqnarray}\label{eq11}
\frac{\partial^2 \delta \psi_{\omega}}{\partial x^2} + \frac{\partial^2 \delta \psi_{\omega}}{\partial y^2}= \frac{s^2}{\omega(\omega + i\nu_n)}\frac{\partial^2 \delta \psi_{\omega}}{\partial x^2}\cdot \delta(y),
\end{eqnarray}
where $\delta (y)$ is the Dirac delta-function describing  thinness of the G-channel.
Solving Eq.~(11) considering that $\delta \psi_{\omega}(x,y) = \delta \varphi_{\omega}(x)
\exp(- \gamma_{\omega}^{GLD}|y|/l_n)$, and accounting for the specifics of the  current induced in the blade-like conducting electrodes~\cite{39}, for the n-region admittance we obtain~\cite{30}

\begin{eqnarray}\label{eq12}
Y_{\omega} = i\omega\biggl[\frac{\kappa}{4\pi}\frac{J_0(\gamma_{\omega}^{GLD})}{\sin (\gamma_{\omega}^{GLD})}
-C_l\biggr].
\end{eqnarray}
Here
\begin{eqnarray}\label{eq13}
\gamma_{\omega}^{GLD} = \frac {\hbar^2\kappa\,l_n\omega(\omega+ i\nu_n)}{4e^2\mu_n} =\frac{\pi\omega(\omega + i\nu_n)}{(\Omega^{GLD})^2}, 
\end{eqnarray}

\begin{eqnarray}\label{eq14}
 \Omega^{GLD} = \frac{e}{\hbar}\sqrt{\frac{4\pi\mu_n}{\kappa\,l_n}} \propto \sqrt{\frac{1}{l_n}},
\end{eqnarray}
$J_0(\xi)$ is the Bessel function of the first kind, and  $C_l$ is the source-drain geometrical capacitance
in the structures with the blade-like side contacts per unit of the GLD width. The latter, accounting for the blade-like shape of the contact is equal to $C_l \simeq (\kappa/4\pi) {\cal L}_l$ with 
${\cal L}_l \sim 1$ being a  logarithmic factor~\cite{39} (see also~\cite{40,41,42}).

Considering 
that 

\begin{eqnarray}\label{eq15}
 \delta J_{\omega}^{BE} = b\Lambda \delta J_{\omega}^{BE} + Y_{\omega} \biggl(\delta V_{\omega} - \frac{l_i}{\sigma_i}\delta J_{\omega}^{BE}\biggr), 
\end{eqnarray}
for the GLD impedance  $Z_{\omega}^{GLD}$ we arrive at the following expression:

\begin{eqnarray}\label{eq16}
 \frac{Z_{\omega}^{GLD}}{R_i} = \frac{(1- b\Lambda)}{R_iY_{\omega}} + 1 + \rho_A\nonumber\\
 = - i\frac{(1- b\Lambda)}{\eta
 \displaystyle\biggl[\displaystyle\frac{J_0(\gamma_{\omega}^{GLD})}{\sin (\gamma_{\omega}^{GLD})}
- 1\biggr]}\biggl[\frac{(\Omega^{GLD})^2}{\pi\omega\nu_n}\biggr] + 1 + \rho_A,
\end{eqnarray}
where $t_{i} = R_iC_l \simeq (\kappa\,R_i/2\pi)  {\cal L}_l= l_i/v_W$.
When $b\Lambda$ and $\omega$ tend to zero, $Y_{\omega}$ and $Z_{\omega}^{GLD}$ tend to  $Z_0^{GLD} =R_n + R_i + R_A$. In the high-frequency limit, $Z_{\omega}^{GD} \simeq R_i + R_A + i/\omega\,C_l \simeq R_i+R_A$.

Figure~4 shows the spectral characteristics of the GLD impedance calculated using Eq.~(16) for GLDs
with different resonant plasma frequencies (due to different $k$) 
at different value of the drag parameter $b\Lambda$ ($b\Lambda > 1$)
assuming that  $\mu_n =25$~meV, $l_n = 1.0~\mu$m, $l_n/\l_i = 10$, 
$\kappa = 4 - 6$, $\eta =6.1 - 9.2$,  ${\overline \nu}_n = 0.75$~ps$^{-1}$, $\rho_A = 1$, and $h = 1000$~cm$^2$/s. 
The n-region length is chosen to be as twice as larger  than that in Fig.~3 related to  the GFETs.
Nevertheless, the values of GLD characteristic  plasma frequency $\Omega^{GLD}$ are larger than the frequency $\Omega^{GFET}$ due to the difference of these frequencies dependences on $l_n$ and $d$. It also assumed that in GLD the viscosity is four times larger to provide the same value of ${\overline \nu}_n$ as for the GLD. 

As seen from Fig.~4, at sufficiently large $b\Lambda$ the impedance real part Re~$Z_{\omega}^{GLD}$ exhibits  deep minima at certain frequencies.
In the first minima Re~$Z_{\omega}^{GLD} < 0$. The second minimum Re~$Z_{\omega}^{GLD}$ is small
being, nevertheless, positive. The latter is attributed to a relatively strong plasma oscillation damping due to the viscosity effect.
The frequency dependences, including the magnitudes in the minima, shown in Figs.~4(a) and 4(b) appears to be rather similar, except for the values of the frequencies $\omega_1/2\pi$ and  $\omega_2/2\pi$ corresponding 
to the Re~$Z_{\omega}^{GLD}$ minima.
These frequencies,  $\omega_1/2\pi \simeq 1.5$~THz and $\omega_2/2\pi \simeq 2.9$~THz [for the GLD with $\Omega^{GLD}/2\pi = 2.71$~THz, see Fig.~4(a)]
and to the minima $\omega_1/2\pi \simeq 1.2$~THz and $\omega_2/2\pi \simeq 2.35$~THz [for the GLD with $\Omega^{GLD}/2\pi = 2.21$~THz, see Fig.~4(b)], are smaller than the pertinent values of the characteristic frequencies 
$\Omega^{GLD}/2\pi$ and $3\Omega^{GLD}/2\pi$.   
This can be explained by a substantial contribution of the GLD geometrical capacitance to the plasma resonances. Such a capacitance increases the net capacitance in comparison with the electron capacitance of the n-region (which determines $\Omega^{GLD}$).
As in the GFETs, the impedance imaginary part Im~$Z_{\omega}^{GLD}$ changes its sign at the frequencies corresponding to the real part minima. Generally, the spectral characteristics of $Z_{\omega}^{GLD}$ are qualitatively similar
to those of the GFETs. 
However, there are the following distinctions: (a) the GLD characteristic and the resonant frequencies are larger
than those of the GFETs for the same n-region length, (b) the resonant frequencies in the GLDs are markedly smaller than the characteristic frequencies (while in the GFETs these frequencies are rather close to each other), and (c) the resonant peaks in the GLDs are affected by the viscosity effect  for a lesser degree. 
Indeed, the plots in Fig.~4, are akin to the plots in Fig.~3, although the former ones exhibit deeper minima
despite plotted for
 a much larger viscosity ($h = 1000$~cm$^2$/s compared with $h = 250$~cm$^2$/s used for Fig.~3). The point is that the wave numbers of the plasma oscillations in the GLDs
are smaller than in the GFETs at the same frequencies. 

However, as shown in the inset in Fig.~4(b),  Re~$Z_{\omega}^{GLD}$
can be negative at the frequency of the second plasma resonance $\omega_2/2\pi \simeq 2.35$~THz at a smaller, but realistic viscosity values.

Generally, the specifics of the $Z_{\omega}^{GLD}$ spectral characteristics implies that the GLD can exhibit the plasma instability analogously to the GFETs.

\section*{4. Plasma instability}

We focus below of the instability in the GFETs 
since the spectral characteristics of the GFET and GLD impedance are qualitatively similar (negative real part and changing sign imaginary part), both corresponding to the possibility of the plasma instability,

To find the conditions of the plasma instability in the GFETs, the frequency of the self-excited modes
$\omega^{\prime}$, and their growth rate $\omega^{\prime\prime}$ corresponding to the plasma instability in the GFETs, 
we use the dispersion equation governing the plasma oscillation in the following form:
 $Z^{GFET}_{\omega^{\prime} + i\omega^{\prime\prime}} = 0$. Invoking Eq.~(7) and setting 
 Re~$Z^{GFET}_{\omega{\prime} + i\omega^{\prime\prime}} = 0$ and
Im~$Z^{GFET}_{\omega{\prime} + i\omega^{\prime\prime}} = 0$, for the growth rate of the  fundamental plasma mode we find


\begin{eqnarray}\label{eq17}
\omega^{\prime\prime} \simeq -\frac{\nu_n}{2}\biggl[1 +\frac{8}{\pi^2}\frac{(1-b\Lambda)}{\eta(1+ \rho_A)}\biggl(\frac{\Omega^{GFET}}{\nu_n}\biggr)^2\biggr]. 
\end{eqnarray}
If $b\Lambda < 1$ (no electron drag or a weak electron drag), Eq.~(17) yields 
$\omega^{\prime\prime} < 0$. This corresponds to the damping of the plasma oscillations with the frequency 
$\simeq \Omega^{GFET}$. At $b\Lambda = 0$, i.e., in the absence of the electron drag, and 
$R_A = 0$, Eq.~(17) yields
$\omega^{\prime\prime} = -\nu_n/2 +t_{RC}^{-1}$, where $t_{RC} = R_iC_n$ is the time of the gated the n-region recharging via the i-region (with the resistance $R_i$, see above).

However, if

\begin{eqnarray}\label{eq18}
b\Lambda  > 1 + \frac{\pi^2\eta(1+ \rho_A)}{8}\biggl(\frac{\nu_n}{\Omega^{GFET}}\biggr)^2,
\end{eqnarray}
the amplitude of the plasma oscillations  is growing, i.e., the plasma instability takes place. 
When $\nu_n \ll \Omega^{GFET}$, inequality~(19) can be satisfied at  
much  smaller $b\Lambda$ than 
required for the GFET source-drain characteristics of the S-type~\cite{17}. 
Considering that  $b\Lambda \simeq 4\pi\,el_i J_0^{BE}/\kappa\,v_W\mu_n$~\cite{17}, inequality~(18) can be presented as

\begin{eqnarray}\label{eq19}
J_{0}^{BE} > J_{th}^{BE} = {\overline J}_{th}^{BE} \biggl[1 + \frac{\pi^2\eta(1+ \rho_A)}{8}\biggl(\frac{\nu_n}{\Omega^{GFET}}\biggr)^2\biggr],\qquad
\end{eqnarray}
where 
\begin{eqnarray}\label{eq20}
{\overline J}_{th}^{BE} = \biggl(\frac{\kappa\,v_W\mu_n}{4\pi\,el_i}\biggr). 
\end{eqnarray}
For typical parameters $\kappa = 4-6$, $l_i = 0.1~\mu$m,  
 $l_n = 0.5~\mu$m, $d = 5\times 10^{-6}$~cm, $\mu_n = 25$~meV, $\nu_n = 1$~ps$^{-1}$, and $\rho_A = 1$, we arrive at the following estimates: $\eta \simeq 6 - 9$, $\Omega^{GFET}/2\pi = 0.87 - 1.08 $~THz, and $J_{th}^{BE} \simeq (130 -190) $~mA/cm. The latter values of the threshold dc current             
  are markedly smaller than that at which the emission of optical phonons by the BEs (affecting the GFET characteristics) becomes essential [$J_0 =\kappa\,v_W \hbar\omega_0/2\pi\,el_i \simeq (1410 - 2115)$~mA/cm, where $\hbar\omega_0 \simeq 200$~meV is the optical phonon energy~\cite{16,18}].

The electron viscosity leads to a strong damping of the higher plasma oscillation harmonics. This 
results in larger values of $b\Lambda$ and $J_{th}^{BE}$ (because of a larger $\nu_n$) and 
complicates (or prevents) the appearance  
of the plasma instability with the self-excitation of these harmonics.

\section*{5. Comments}

\subsection*{Distributed model versus lumped-element model}

Compare the GFET impedance calculated using the lumped capacitance and inductance model with
that obtained above. The latter is given by (in the present notations)~\cite{18}

\begin{eqnarray}\label{eq21}
\frac{Z_{\omega}}{R_i} = \frac{(1 -b\Lambda)}{\eta                                                                                                                                                                                       (1 +i {\cal F}_{\omega})}\frac{(\omega^2 + \nu_n^2)}{\nu_n^2}
  +1 + \rho_A,
\end{eqnarray}
 where ${\cal F}_{\omega} = (\omega/\nu_n)[(\Omega_l^{GFET})^2 - \omega^2 - \nu_n^2]/(\Omega_l^{GFET})^2$ with $\Omega_l^{GFET}$, which differs from $\Omega^{GFET}$ given by Eq.~(6) by a factor of $\sqrt{2/\pi}$. This is because the distributed model accounts for a deviation of the ac potential distribution  in the n-region from a linear distribution (in contrast to the lumped-element model). Equation~(21) yields the resonant peak height equal to the fundamental resonant peak ($m=1$) height
 described by Eq.~(10). However, the main distinction in the spectral characteristics of the GFET impedance calculated by the distributed and lumped-element models is that the former can describe  a multiple peak structure.

\subsection*{Effect of the side contacts shape on the plasma resonances in GLDs}

  Equations~(12) - (14)
 correspond to the case of the GLDs with the blade-like side contacts. If the thickness of the side contacts in the GLDs, 
 $D > l_i + l_n \simeq l_n$ (the GLDs with bulk contacts),  
we have to account  
  for the features of the displacement  current induced by the carriers 
  between the source and drain (bulk) contacts. In this case,   for the n-region admittance we obtain [compare with Eq.~(12)] 
 
\begin{eqnarray}\label{eq22}
 Y_{\omega} = i \biggl(\frac{\sigma_n\nu_n}{\omega + i\nu_n} - \omega\,C_b\biggr).
\end{eqnarray}
We estimate  the capacitance $C_b$  as $C_b \sim (\kappa/4\pi){\cal L}_b $ with ${\cal L}_b  \sim D/l_n$.
 In the case under consideration, Eqs.~(15) and (22) lead to the following formula for $Z_{\omega}^{GLD}$:
 
\begin{eqnarray}\label{eq23}
 \frac{Z_{\omega}^{GLD}}{R_i} 
 = - i\frac{(1- b\Lambda)}{\displaystyle\frac{\omega\,t_i{\cal L}_b}{2}\biggl[\displaystyle\frac{4\pi\,\sigma_n\nu_n}{ {\cal L}_b\kappa\,l_n\omega(\omega + i\nu_n)} - 1\biggr]} + 1 + \rho_A\nonumber\\
 =- i\frac{(1- b\Lambda)}{\displaystyle\frac{\omega\,t_i{\cal L}_b}{2}\biggl[\displaystyle\frac{(\Omega_b^{GLD})^2}{\omega(\omega + i\nu_n)} -1\biggr]} + 1 + \rho_A,\qquad
\end{eqnarray}
 Hence $t_i{\cal L}_b \simeq (l_i/v_W)(D/l_n) = t_i(D/l_n)$ and the plasma frequency in the GDs with bulk contacts is given by

\begin{eqnarray}\label{eq24}
\Omega_b^{GLD} \simeq \frac{e}{\hbar}\sqrt{\frac{4\mu_n}{\kappa\,l_n{\cal L}_b}} \sim \frac{e}{\hbar}\sqrt{\frac{4\mu_n}{\kappa\,D}}.
\end{eqnarray} 
 Equation~(24) can be presented in the form of Eq.~(21) with ${\cal F}_{\omega}$ including $\Omega_b^{GLD}$ instead of $\Omega_l^{GFET}$ with $\Omega_b^{GLD}/\Omega_l^{GFET} = l_n/\sqrt{2dD}$ (i.e., 
 $\Omega_b^{GLD}/\Omega^{GFET} = 2l_n/\pi\sqrt{dD}$). Thus, our model for the GLDs with relatively thick contacts
 yields the results similar to those obtained in the framework of the GFET lumped-element model, corresponding,
 in particular, to a single plasmonic resonance. However, the resonant plasma frequencies are different due to a difference of the geometrical gate capacitance in the GFET and the source-drain geometric capacitance of  the GLDs with bulk contacts.

\subsection*{Output THz power}

To estimate the maximum output THz power, one can determine the 
variation $\delta \Lambda$ of the parameter $\Lambda$ and hence the 
swing of the dc bias current $\Delta J_0^{BE}$ corresponding to Re~$Z_{\omega}^{GFET} < 0$ (or Re~$Z_{\omega}^{GLD} < 0$ ). From Eqs.~(7) we find 

\begin{eqnarray}\label{eq25}
b \Delta \simeq (b\Lambda - 1) + \eta(\nu_n/\Omega^{GFET})^2 (1+ \rho_A).
\end{eqnarray} 
At the parameters used for Fig.~3(a), we obtain $\eta(\nu_n\Omega^{GFET})^2  \simeq 0.2$.
Considering that at $J_0^{BE} \leq J_0$ (we disregard the case or relatively high dc bias current
at which the optical phonon emission markedly affects the GFET (GLD) characteristics), 
 $\Delta \Lambda = 2\Delta J_0^{BE}/J_0$, for $b\Lambda = 2$ and  $b\Lambda = 3$ [as in Fig.~3(a)],
from Eq.~(25)  we obtain 
$\Delta J_0^{BE}/J_0 \simeq 0.6 - 0.7$. Considering that at the above parameters $J_0 \simeq 1.4$~A/cm and for the GFET width $H = (10 -14)~\mu$m $R_i 
= R_A \simeq 100 - 140$~Ohm,  one can obtain $\Delta J_0^{BE} \simeq (0.85 -1.4)$~mA.
In this case the maximum output THz power $P_{\omega}^{GFET}$ at the frequency $\omega/2\pi = 1$~THz
is estimated as $P_{\omega} \simeq 100 -200~\mu$W. Similar estimates can be obtained for the GDLs.

\section*{6. Conclusions}

Our calculations of the frequency dependendencies   of the n$^+$-i-n-n$^+$  and GFETs and GLDs impedances  accounting for the resonant response
of the electron plasma in the n-regions, damping of the plasma oscillations due to the electron viscosity, and the Coulomb drag of the QEs by the injected BEs
show that the impedance real part Re~$Z_{\omega}$ in both  GFETs and GLDs can be negative if the drag effect is sufficiently strong. Since in the frequency range, where  Re $Z_{\omega} <0$, Im~$Z_{\omega}$ changes  sign, i.e., turns zero at a certain THz signal frequency, the electron plasma can be unstable toward the self-excitation of the plasma oscillations (the effect of the plasma instability). This can enable the emission of the THz radiation. The electron viscosity can effectively suppress the higher plasma resonances, although this effect is weaker in the GLDs in comparison with the GFETs.
The results related to the GFETs confirm the predictions obtained using a simplified model of the gated n-region  as a plasmonic cavity except not accounting for the possibility of the plasma frequency harmonics self-excitation.
We demonstrated that the GLDs with the ungated n-region formed by chemical selective doping can also exhibit
the plasma instability and generation of the THz radiation. In these devices, the plasma resonant frequency and,
hence, the frequency of the emitted THz radiation, can markedly exceed that in the GFETs for the same n-region length.  
The obtained results imply that the n$^+$-i-n-n$^+$ GFETs and GLDs can be used in novel THz radiation sources. A  similar instability could also occur 
in p$^+$-i-p-p$^+$ (including the structures  based on G-multilayers with the carriers induced by the gate voltage~\cite{43} or doping) and the p$^+$-p-i-n-n$^+$ single G-layer (i.e., GTTs~\cite{20})  or G-multilayer~\cite{44}   structures
with the Zener-Klein interband tunneling generation of ballistic carriers.

\section*{Acknowledgments}
The work at RIEC and UoA was supported by Japan Society for Promotion of Science (KAKENHI Grants No. 21H04546 and No. 20K20349), Japan;  A01No. H31/A01.  The work at RPI was supported by Office of Naval Research (N000141712976, Project Monitor Dr. Paul Maki).


 \end{document}